\documentclass[prb,10pt,amsmath,amssymb,letterpaper,nobalancelastpage,aps,showpacs,preprintnumbers,twocolumn]{revtex4-1}
\DeclareMathOperator{\fracpart}{frac}
\providecommand{\ket}[1]{| #1 \rangle}
\providecommand{\expup}[1]{\text{e}^{#1}}
\providecommand{\rmi}[0]{\text{i}}
\providecommand{\rmd}[0]{\text{d}}

\providecommand{\abs}[1]{\bigl| #1 \bigr|}
\def\fig#1{Fig.~\ref{#1}}
\def\vec#1{\text{\bfseries#1}}
\def\mat#1{\text{\sffamily\bfseries#1}}

\usepackage{graphicx,color,bm}

\begin{document}
\title{Quantum phase transition in a multicomponent anyonic Lieb-Liniger model}
\author{Raul A. Santos}
\affiliation{C. N. Yang Institute for Theoretical Physics, State University of New York at Stony Brook, New York 11794-3840, USA}
\author{Francis N. C. Paraan}
\affiliation{C. N. Yang Institute for Theoretical Physics, State University of New York at Stony Brook, New York 11794-3840, USA}
\author{Vladimir E. Korepin}
\affiliation{C. N. Yang Institute for Theoretical Physics, State University of New York at Stony Brook, New York 11794-3840, USA}
\date{\today}

\begin{abstract}
We study a one-dimensional multicomponent anyon model that reduces to a multicomponent Lieb-Liniger gas of impenetrable bosons (Tonks-Girardeau gas) for vanishing statistics parameter. At fixed component densities, the coordinate Bethe ansatz gives a family of quantum phase transitions at special values of the statistics parameter. We show that the ground state energy changes extensively between different phases. Special regimes are studied and a general classification for the transition points is given. An interpretation in terms of statistics of composite particles is proposed. 
\end{abstract}

\pacs{05.30.Pr, 02.30.Ik, 73.43.Nq}

\preprint{YITP-SB-12-10}

\maketitle
\hbadness=10000


{\em Introduction---}
Anyons play an important role in topological quantum computation.\cite{kitaev2003,nayak2008} They are the natural quasiparticle excitations of fractional quantum Hall states.\cite{arovas1984} While the fractional quantum Hall effect occurs in a two-dimensional electron liquid, its edge excitations are described by one-dimensional anyons.\cite{halperin1982a,wen1990a} These edge states are essential for performing quantum computation operations by topological quantum gates.\cite{averin2002,dassarma2005a}

A model for one-dimensional anyons is the Lieb-Liniger model\cite{lieb1963} with fractional exchange statistics.\cite{kundu1999} An important feature of this model is that it is solvable by the Bethe ansatz.\cite{batchelor2007} In this model the statistics parameter (exchange phase) does not modify the bulk ground state energy in large systems (correction terms vanish in the thermodynamic limit).\cite{batchelor2006a} However, the effect of anyon statistics on other nonlocal physical quantities are more prominent. For example, momentum distributions, reduced density matrices, and other correlation functions have a nontrivial dependence on the statistics parameter.\cite{batchelor2006a,batchelor2006b,patu1,*patu2,*patu3,*patu4,patu0,patu00,patu000,hao2008,calabrese2007a,santachiara2007,santachiara2008,calabrese2009,hao2009} Experiments involving photons\cite{matthews2011,vanexter2012} and ultracold atoms\cite{keilmann2011} that have anyonic exchange statistics in one-dimension are accessible with current technology.

In this paper, we generalize the anyonic Lieb-Liniger model to the multicomponent case. As known for the integrable multicomponent Bose gas,\cite{kulish1981b,pu1987,izergin1989} the additional internal degrees of freedom lead to diverse physical effects.\cite{gu2002,fuchs2005,batchelor2006c,guan2007,deuretzbacher2008,kleine2008a,*kleine2008b,caux2009,kolezhuk2010,klumperpatu2011,klauser2011} In particular, we consider a gas of two mutually impenetrable anyonic species. This exactly solvable model may prove useful in studies of the edge states of bilayer fractional quantum Hall systems.\cite{halperin1983a,yang1994a,lopez2001a,mazo2011a}  We encounter an interesting quantum phase transition\cite{sachdevqpt} (QPT) that is controlled by the statistics parameters of the model. This QPT distinguishes between two phases with ground states that are described by either one or two Dirac seas (one-dimensional Fermi spheres). The ground state energies of these phases can therefore differ significantly by an extensive amount.

We first motivate this work by describing a similar QPT that is controlled by the dynamical interparticle coupling in a multicomponent boson gas. We then proceed with a formal description of our multicomponent anyonic model. This model is formulated in terms of quantum fields satisfying generalized commutation relations. Next, we calculate exact eigenfunctions and the energy spectrum of this theory. We finally show how the statistics parameters determine whether the ground state is described by one or more Dirac seas. This result allows us to construct a phase diagram for the ground state in the space of all possible statistics parameters.

{\em Bosons---}
Let us consider a generalized two-component boson gas in the impenetrable or Tonks-Girardeau (TG) limit.\cite{girardeau1960} Hereafter, we sometimes use a pseudospin picture and refer to these two components as different spin-1/2 projections. The relevant quantum Hamiltonian is a multicomponent version of the Lieb-Liniger model
\begin{multline}\label{bosonh}
	\mathcal{H}_\text{B} = \sum_{\alpha=1}^{2}\int_{0}^L\bigl[\partial_x\phi_{\alpha}^\dag(x)\bigr]\bigl[\partial_x\phi_{\alpha}(x)\bigr]\\ + \sum_{\beta=1}^2c_{\alpha\beta}\,\phi_{\alpha}^\dag(x)\phi_{\beta}^\dag(x)\phi_{\beta}(x)\phi_{\alpha}(x)\,\rmd x.
\end{multline}
The fields $\phi_\alpha(x)$ are independent canonical Bose fields satisfying periodic boundary conditions in a box of length $L$. The field index $\alpha\in\{1,2\}$ labels different particle species. Energy eigenstates of this Hamiltonian are simultaneous eigenstates of the number operator \smash{$\mathcal{N}_\alpha = \int_0^L \phi_\alpha^\dag(x)\phi_\alpha(x)\,\rmd x$}. Thus, the number of type-$\alpha$ particles $N_\alpha$ are good quantum numbers. 

We look at two extreme limits of the interspecies coupling. One is the decoupled limit in which the matrix $\mat{c} = (\begin{smallmatrix} c_{11} & c_{12} \\c_{21} &c_{22}\end{smallmatrix})$ is diagonal. That is, $c_{12}=c_{21}=0$ while both $c_{11}$ and $c_{22}$ tend to infinity. This limit describes two non-interacting TG gases. The quantum mechanical eigenfunctions of the model are Bethe wavefunctions. The impenetrable nature of the gas implies that this wavefunction vanishes when the coordinates of particles of the same spin coincide. Thus, the wavefunction $\chi_{\text{B}}$ is a product of two symmetrized Slater determinants:
\begin{equation}\label{bosontwosea}
	\chi_{\text{B}}^{\text{II}}(\vec{z}_1,\vec{z}_2|{\bm{\lambda}_{1}},{\bm{\lambda}_{2}}) = \frac{\abs{\det e^{\rmi\lambda_{1,j}z_{1,k}}}}{\sqrt{\smash[b]{N_1!}}}\times \frac{\abs{\det e^{\rmi\lambda_{2,m}z_{2,n}}}}{\sqrt{\smash[b]{N_2!}}}.
\end{equation} 
Each $N_\alpha\times N_\alpha$ determinant depends only on the coordinates $\vec{z}_{\alpha} \equiv \{z_{\alpha,i}\}$ of type-$\alpha$ particles, $i\in\{1,\dotsc,N_\alpha\}$. They are characterized by independent sets of $N_\alpha$ spectral parameters or momenta ${\bm\lambda}_\alpha\equiv\{\lambda_{\alpha,i}\}$. The momenta within each set are distinct, otherwise the wavefunction \eqref{bosontwosea} vanishes (Pauli exclusion).\cite{izergin1982a} Hence, the ground state is described by two Dirac seas with Fermi levels $q_\alpha = (N_\alpha -1)\pi/L$. In the thermodynamic limit where particle numbers $N_\alpha$ tend to infinity at fixed linear density $D_\alpha$, the ground state energy per unit length is $E_0^\text{II}/L = (D_1^3 + D_2^3)\pi^2/3$.

The opposite limit is of mutual repulsion where all elements of $\mat{c}$ tend to infinity.\cite{guan2007} In this example both components are coupled dynamically. The eigenfunctions in this case are 
\begin{equation}\label{bosononesea}
\chi_{B}^\text{I}(\vec{z},{\bm{\sigma}_\vec{N}}|\bm{\lambda})=A(\bm\sigma_{\vec{N}})\times\frac{\abs{\det e^{\rmi\lambda_{j}z_{k}}}}{\sqrt{N!}}.
\end{equation}
In this formula, the vectors $\vec{z}$ and $\bm\sigma_\vec{N}$ give the position $z_i$ and spin $\sigma_i$ of the $i\textsuperscript{th}$ boson. The vector ${\bm{\sigma}}_{\vec{N}}$ has $N_\alpha$ components equal to $\alpha$. The factor $A(\bm\sigma_{\vec{N}})$ gives the overall symmetry of the wavefunction under particle exchanges. It is equal to $+1$ when coordinates of particles with identical spins are interchanged. Otherwise, when coordinates of different spins are exchanged it is equal to $+1$ or $-1$. A choice of $\pm 1$ will correspond to a compatible irreducible representation (Young tableau) of the symmetric group $S_2$.\cite{yang1967,sutherland1968a} Meanwhile, the wavefunction \eqref{bosononesea} is characterized by a single set $\bm\lambda\equiv\{\lambda_j\}$ of $N$ distinct momenta. The ground state of this two-component gas is thus described by a single Dirac sea with Fermi level $q = (N-1)\pi/L$.\footnote{In general, the boundary conditions satisfied by the many-boson wavefunction for this result to hold depend on the parity of $N_1$ and $N_2$. For example, if the antisymmetric Young tableau is chosen and $N_1$ and $N_2$ have different parities, the wavefunction must satisfy mixed periodic and antiperiodic boundary conditions.} In the thermodynamic limit with total particle density $D = D_1 +D_2$, the ground state energy per unit length is $E_0^\text{I}/L = D^3\pi^2/3$. This value is equal to the ground state energy density of a single component TG gas and fully-polarized free Fermi gas with the same total particle number.\cite{guan2007,yangyou2011}

The ground states of these boson gases are fundamentally different. One is characterized by a single Dirac sea and the other by two Dirac seas. The ground state energy of the mutually repulsive gas is much higher than the non-interacting example because $(D_1+D_2)^3>D_1^3 + D_2^3$ (equal only in the ``ferromagnetic'' case when one of $D_\alpha = 0$). Restructuring of the ground state between these two limits is controlled dynamically, that is, through the interspecies coupling coefficients. In the following section we seek to find a similar transition that is generated statistically by generalized exchange phases between particles. 

{\em Anyons---}
Our one-dimensional anyonic model consists of two anyon fields $\Psi_{\alpha}(x)$ with $\alpha\in\{1,2\}$. These fields satisfy the exchange relations (no sums involved)
\begin{align}\label{anyonicSYM}\nonumber
 \Psi_{\alpha}(x)\Psi_{\beta}(y)&=U_{\alpha\beta}(x-y)\Psi_{\beta}(y)\Psi_{\alpha}(x),\\
 \Psi_{\alpha}^\dag(x)\Psi_{\beta}^\dag(y)&=U_{\alpha\beta}(x-y)\Psi_{\beta}^\dag(y)\Psi_{\alpha}^\dag(x),\\\nonumber
 \Psi_{\alpha}(x)\Psi_{\beta}^\dag(y) &=U_{\alpha\beta}^*(x-y)\Psi_{\beta}^\dag(y)\Psi_{\alpha}(x) + \delta_{\alpha\beta}\delta(x-y),
\end{align}

\noindent where	$U_{\alpha\beta}(x) \equiv \expup{2\pi\text{i}\kappa_{\alpha\beta}\epsilon(x)}$. Here $\bm{\kappa} = (\begin{smallmatrix} \kappa_{11} & \kappa_{12}\\\kappa_{21}&\kappa_{22}\end{smallmatrix})$ is a real symmetric matrix and $\epsilon(x)$ is the sign of $x$ with $\epsilon(0)=0$. All cases of interest are contained in the parameter space $\kappa_{\alpha\beta}\in[0,1]$. The exchange relations \eqref{anyonicSYM} therefore depend on continuous parameters that interpolate between commuting (bosonic) and anticommuting (fermionic) fields at different points in space and equal times. At the same space-time point we can choose to recover either bosonic fields or fermionic fields.\cite{girardeau2006a} The commutation relations here reduce to bosonic ones at the same point.

The exchange phase of a localized bunch of $N$ anyons can be computed from these exchange relations. Let $\Phi^\dag(Z)$ be a composite operator \smash{
$\Phi^\dag(Z)\equiv \prod_{i=1}^{N}\Psi_{\sigma_i}^\dag(z_i)$}
for some configuration of anyons with positions $\vec{z}$ and spins $\bm\sigma_\vec{N}$. Further, let all $\{z_i\}$ be in the neighborhood of $Z$ and all $\{y_i\}$ be in the neighborhood of $Y$ so that all $z_i > y_j$ for some $Z>Y$. Hence, the exchange relation satisfied by a composite particle resulting from the fusion of $N$ anyons is, for $Z>Y$,
\begin{equation}\label{anyonfusion}
 \Phi^\dag(Z)\Phi^\dag(Y)=\expup{2\pi \rmi(\kappa_{11}N_1^2+2\kappa_{12}N_1N_2+\kappa_{22}N_2^2)}\Phi^\dag(Y)\Phi^\dag(Z).
\end{equation}

Now, the quantum Hamiltonian we consider is an anyonic version of the mutually repulsive boson model \eqref{bosonh} with all $c_{\alpha\beta} \to \infty$. It is given by

\begin{multline}\label{anyonh}
\mathcal{H} = \sum_{\alpha=1}^{2}\int_{0}^L\bigl[\partial_x\Psi_{\alpha}^\dag(x)\bigr]\bigl[\partial_x\Psi_{\alpha}(x)\bigr]\\ + c\sum_{\beta=1}^2\Psi_{\alpha}^\dag(x)\Psi_{\beta}^\dag(x)\Psi_{\beta}(x)\Psi_{\alpha}(x)\,\rmd x.
\end{multline}

\noindent where $c$ is a coupling constant that we will send to $+\infty$. In this section, we solve for the eigenfunctions and spectrum of this model by the coordinate Bethe ansatz.

The component number density operators $\rho_\alpha(x) \equiv \Psi_\alpha^\dag(x)\Psi_\alpha(x)$ obey the usual commutation relations
\begin{align}
 [\Psi_\alpha(x),\rho_\beta(y)] &= \delta_{\alpha\beta}\delta(x-y) \Psi_\beta(y), \nonumber \\
 [\Psi_\alpha^\dag(x),\rho_\beta(y)] &= -\delta_{\alpha\beta}\delta(x-y) \Psi_\beta^\dag(y). \label{crnum}
\end{align}
We find that the numbers $N_\alpha$ of type-$\alpha$ anyons are individually conserved.

We stress that this anyonic model \eqref{anyonh} is related, but not equivalent, to the bosonic model \eqref{bosonh}. To see this, we use a generalized Jordan-Wigner transformation
\begin{align}
	\phi_\alpha(x) &= \expup{2\pi\text{i}\sum_\gamma\kappa_{\alpha\gamma}\int_0^x \rho_\gamma(z)\,\text{d}z} \Psi_\alpha(x), 
	\nonumber \\
	\phi_\beta^\dag(y) &= \Psi_\beta^\dag(y)\expup{-2\pi\text{i}\sum_\gamma\kappa_{\beta\gamma}\int_0^y \rho_\gamma(z)\,\text{d}z},\label{jw}
\end{align}
to obtain independent canonical Bose fields $\phi_\alpha(x)$. Applying the inverse transformation to the anyonic model \eqref{anyonh} results in a gauge-transformed Lieb-Liniger model
\begin{multline}\label{bosongauge}
	\mathcal{H}_\text{B}'= \sum_{\alpha=1}^{2}\int_{0}^L\bigl[\mathcal{D}_\alpha(x)\phi_{\alpha}(x)\bigr]^\dag\bigl[\mathcal{D}_\alpha(x)\phi_{\alpha}(x)\bigr] \\+ c\sum_{\beta=1}^2\phi_{\alpha}^\dag(x)\phi_{\beta}^\dag(x)\phi_{\beta}(x)\phi_{\alpha}(x)\,\rmd x,
\end{multline}
with covariant derivative $\mathcal{D}_\alpha(x)\equiv \partial_x - 2\pi\rmi\sum_{\gamma}\kappa_{\alpha\gamma}\rho_\gamma(x)$.

Following the treatment of the single component anyonic case,\cite{kundu1999} we map the Hamiltonian (\ref{anyonh}) to a 
many-body quantum mechanical problem. We let the vectors $\vec{z}$ and $\bm\sigma_\vec{N}$ contain the positions $z_i$ and spin $\sigma_i$ of the $i\textsuperscript{th}$ anyon, as done with the mutually repulsive boson gas \eqref{bosononesea}. We introduce the number eigenstates ($N=N_1+N_2$)
\begin{equation}
 \ket{N_1,N_2}=\sum_{\langle\bm\sigma_\vec{N}\rangle}\int\rmd^{N}\negthinspace z \, \chi(\vec{z},{\bm\sigma}_{\vec{N}})\prod_{i=1}^{N}\Psi_{\sigma_i}^\dag(z_i)\ket{0},
\end{equation}
where the anyonic vacuum state satisfies $\Psi_\alpha(z_i)\ket{0} = 0$. The sum over $\langle\bm\sigma_\vec{N}\rangle$ denotes a sum over all configurations with $N_\alpha$ type-$\alpha$ anyons. This construction imposes a certain anyonic symmetry on the wavefunction $\chi$. That is, if $\mathcal{P}_{ij}$ is a particle exchange operator such that $\mathcal{P}_{ij}\chi(\dotsc z_i \dotsc z_j\dotsc,\dotsc \sigma_i \dotsc \sigma_j \dotsc) \equiv \chi(\dotsc z_j \dotsc z_i\dotsc,\dotsc \sigma_j \dotsc \sigma_i \dotsc)$, then
\begin{equation}\label{WFanyonsym}
 \chi=\expup{2\pi\rmi\sum_{l=i+1}^j\kappa_{\sigma_i\sigma_l}\epsilon(z_i-z_l)}\expup{-2\pi\rmi \sum_{l=i+1}^{j-1}\kappa_{\sigma_j\sigma_i}\epsilon(z_j-z_l)} \mathcal{P}_{ij}\chi.
\end{equation}

The associated many-body Hamiltonian acting on the wavefunction $\chi(\vec{z},{\bm\sigma}_{\vec{N}})$ is 

\begin{equation}\label{QMHamiltonian}
 H=-\sum_{i=1}^{N}\partial_{z_{i}}^2+2c\sum_{i<j}^{N}\delta(z_{i}-z_{j}).
\end{equation}

\noindent In the Tonks-Girardeau limit $c\to\infty$ the wavefunction $\chi$ vanishes at the coinciding points $z_{i}=z_j\ (i\ne j)$. 
With the symmetry relations \eqref{WFanyonsym}, we are able to write the anyonic wavefunction $\chi$ as a product of a multicomponent boson wavefunction $\chi_{\text{B}}(\vec{z},{\bm{\sigma}}_\vec{N})$ and a position dependent anyonic phase:

\begin{equation}\label{anyonfactor}
 \chi(\vec{z},{\bm{\sigma}}_\vec{N})=\chi_{\text{B}}(\vec{z},{\bm{\sigma}}_\vec{N})\expup{-\pi\rmi\sum_{j<k}\kappa_{\sigma_j\sigma_k}\epsilon(z_j-z_k)}.
\end{equation}


The symmetry of the anyonic wavefunction $\chi$ between an interchange of anyons of {\em different} spins is not fixed by the anyonic commutation relations (\ref{anyonicSYM}). Like in the bosonic case \eqref{bosononesea}, the wavefunction $\chi_{\text{B}}$ in Eq.~\eqref{anyonfactor} contains a factor corresponding to the appropriate irreducible representation of the symmetric group $S_2$. The choice of Young tableau is related to the particular set of fusion rules that we can assign to the anyons of the model. For simplicity we consider the completely symmetric Young tableau of $N = N_1+N_2$ boxes for the boson wavefunction. That is, we assume that $\chi_\text{B}$ in Eq.~\eqref{anyonfactor} is also symmetric between interchanges of {\em different} types of particles. The antisymmetric case may be obtained from the symmetric case by the transformation $\kappa_{12} \to \kappa_{12} + 1/2$.

With these considerations, the anyonic wavefunction can be explicitly expressed as
\begin{equation}\label{anyonwf}
	\chi(\vec{z},{\bm{\sigma}}_\vec{N}|\bm\lambda)=\frac{\abs{\det \expup{\rmi\lambda_m z_n}}}{\sqrt{N!}}\,\expup{-\pi\rmi\sum_{j<k}\kappa_{\sigma_j\sigma_k}\epsilon(z_j-z_k)},
\end{equation}
where the set of $N$ momenta $\{\lambda_m\}$ are determined by the boundary conditions imposed on $\chi$. We will find it convenient to relabel the momenta $\{\lambda_m\} \to \{\lambda_{\alpha,i}\}$ with Greek index $\alpha\in\{1,2\}$ and Latin index $i\in\{1,\dotsc,N_\alpha\}$. Hence, a state with this wavefunction has total momentum $P = \sum_{\alpha}\sum_{i}\lambda_{\alpha,i}$ and energy $E= \sum_{\alpha}\sum_{i}\lambda_{\alpha,i}^2$.

\begin{figure}[tb]
	\centering
		\includegraphics[width=\linewidth]{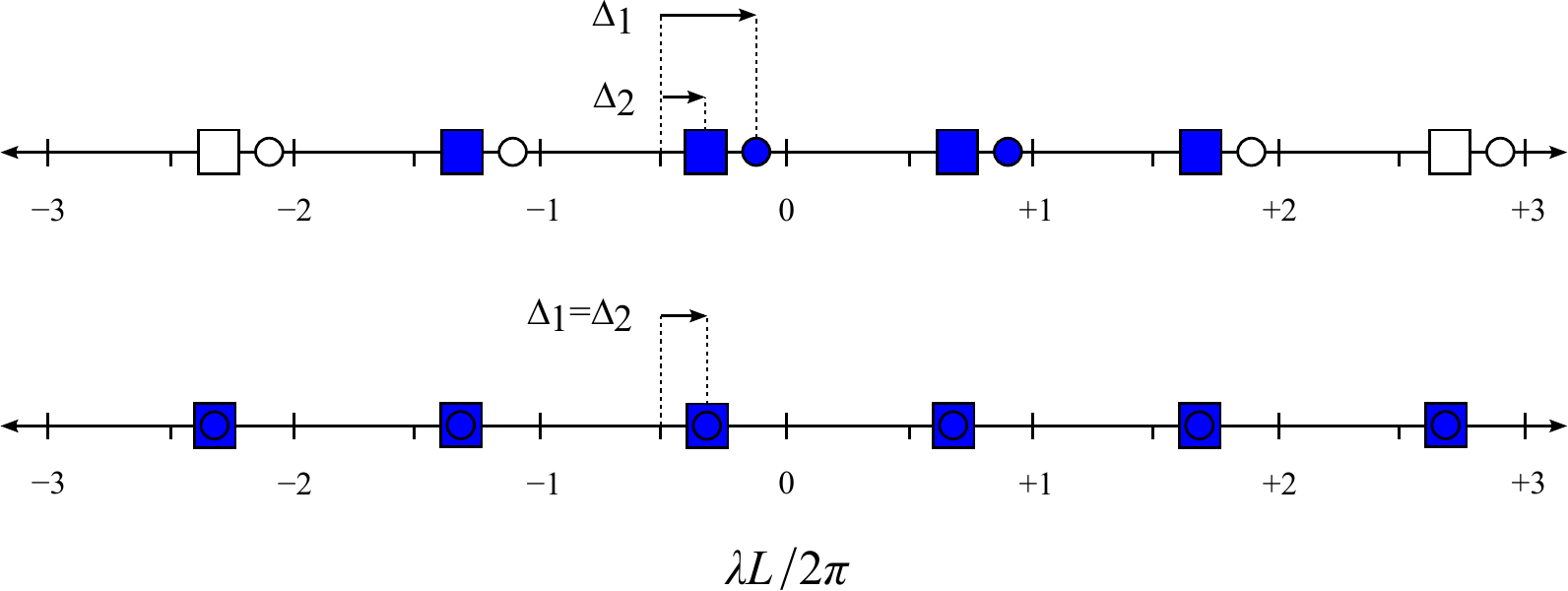}
	\caption{(Color online) Ground state occupation of momentum vacancies for two type-1 and four type-2 impenetrable anyons. Filled vacancies are shaded and the maximum occupancy is one. Squares and circles refer to different sets of vacancies. Each set is shifted from its bosonic limiting value by a statistics-dependent term $\Delta_\alpha$. There are two Dirac seas when the shifts are unequal $\Delta_1\ne\Delta_2$ (top) and one Dirac sea when they are equal $\Delta_1=\Delta_2$ (bottom).}
	\label{fig:fermisphere}
\end{figure}

The boundary conditions on the anyonic wavefunction may be chosen to be periodic or twisted. If we choose the twisted boundary condition to cancel the 
anyonic phase $\expup{-\pi\rmi\sum_{i<j}\kappa_{\sigma_i\sigma_j}\epsilon(z_i-z_j)}$, we recover the energy spectrum of the mutually repulsive boson model. This phenomenon is also observed in the single component case.\cite{patu000} That is, certain twisted boundary conditions on the anyonic wavefunction results in a bosonic wavefunction and vice versa. This correspondence is related to the Jordan-Wigner transformation \eqref{jw} connecting the anyonic model \eqref{anyonh} and gauge-transformed bosonic model \eqref{bosongauge}. 

We now focus on the situation where the anyonic wavefunction $\chi$ satisfies periodic boundary conditions. As discussed for the one component case,\cite{batchelor2007,patu000} periodic boundary conditions on $\chi$ enter in a nontrivial way. Suppressing component labels, the periodic boundary conditions we impose on $\chi(z_1,z_2,\dotsc,z_N)$ are

\begin{align}
 \chi(0,z_2,z_3,\dotsc)&=\chi(L,z_2,z_3,\dotsc) \nonumber\\
 \chi(z_1,0,z_3,\dotsc)&=e^{-4\pi\rmi\kappa_{\sigma_2\sigma_1}}\chi(z_1,L,z_3,\dotsc) \nonumber \\
 \chi(z_1,z_2,0,\dotsc)&=e^{-4\pi\rmi(\kappa_{\sigma_3\sigma_1}+\kappa_{\sigma_3\sigma_2})}\chi(z_1,z_2,L,\dotsc) \nonumber \\
\chi(z_k=0)&=\expup{-4\pi \rmi\sum_{j=1}^{k-1}\kappa_{\sigma_k\sigma_j}}\chi(z_k=L). 
\end{align}

\noindent In logarithmic form, the Bethe equations for the anyonic momenta are therefore

\begin{equation}
	\lambda_{\alpha,i}L = 2\pi {n}_{\alpha,i} + \pi(N-1) + \delta_{\alpha}, \label{bethesoln1}
\end{equation}
where $\{{n}_{\alpha,i}\}$ is a set of integers. Here $\pi(N-1)$ is the accumulated dynamical scattering phase due to the anyon hard cores.
The phase shift $\delta_\alpha$ is
\begin{equation}\label{statshift}
	\delta_\alpha \equiv 2\pi\biggl(- \kappa_{\alpha\alpha} + \sum_{\gamma=1}^2 \kappa_{\alpha\gamma}N_\gamma\biggr).
\end{equation}
This quantity represents the total phase shift due to the exchange statistics. We can also write the Bethe solutions \eqref{bethesoln1} as 
\begin{equation}
	\lambda_{\alpha,i} = \frac{2\pi}{L}( m_{\alpha,i} + \Delta_{\alpha}), \label{bethesoln2}
\end{equation}
where $\{{m}_{\alpha,i}\}$ is a set of half-odd integers for even $N$ (or integers for odd $N$). The shift $\Delta_\alpha$ is defined as
\begin{equation}
	\Delta_{\alpha} \equiv \begin{cases}
\fracpart\Bigl(\dfrac{\delta_\alpha}{2\pi}\Bigr), & \text{if }\fracpart\Bigl(\dfrac{\delta_\alpha}{2\pi}\Bigr)\in [0,1/2], \\
\\
\fracpart\Bigl(\dfrac{\delta_\alpha}{2\pi}\Bigr) - 1, & \text{if }\fracpart\Bigl(\dfrac{\delta_\alpha}{2\pi}\Bigr)\in(1/2,1).
\end{cases}\label{shift}
\end{equation}
where $\fracpart(x)$ is 
$x$ minus the integer part of $x$. With this definition, $\Delta_{\alpha}$ is restricted to the interval $(-1/2,1/2]$.

{\em Quantum phase transition---}
The momenta that label the anyonic wavefunction \eqref{anyonwf} are different from each other. Otherwise the Slater determinant is zero and the wavefunction vanishes. 
Thus, when the shifts \eqref{shift} are equal $\Delta_1=\Delta_2$, the ground state is described by a single filled Dirac sea. That is, $N$ momentum vacancies $\{\lambda_m\}$ with the smallest squares are occupied (see \fig{fig:fermisphere}). On the other hand, when $\Delta_1\ne\Delta_2$ the set of momentum vacancies split into two: $\{\lambda_{1,i}\}$ and $\{\lambda_{2,j}\}$. Two Dirac seas are therefore formed in the ground state. Each is filled with $N_\alpha$ quasiparticles to minimize the ground state energy. Since the shifts $\Delta_\alpha$ depend on the parameters $\kappa_{\alpha\beta}$, we may think of the transition between the single and double Dirac sea ground states as QPTs controlled by $\kappa_{\alpha\beta}$. Hereafter, we will refer to the single Dirac sea phase as degenerate and double sea phase as nondegenerate. 

The ground state of the model generally has nonzero momentum as the anyonic statistics break spatial parity symmetry. In both phases, the ground state acquires a total momentum $P_0=2\pi(D_1 \Delta_1 + D_2\Delta_2)$ for $N_1,N_2$ even. Thus, the statistics parameters modify the ground state energy by a constant (non-extensive) amount. 
However, the ground state energy of these phases are significantly different from each other. The exact ground state energy density for the degenerate phase is
\begin{equation}
	\frac{E_0^\text{I}}{L} = \frac{\pi^2}{3}\biggl[(D_1 + D_2)^3 +\frac{(12\Delta^2-1)(D_1 + D_2)}{L^2} \biggr],
\end{equation}
 where $\Delta\equiv\Delta_1=\Delta_2$. Meanwhile, for the nondegenerate phase it is equal to 
\begin{equation}
	\frac{E_0^\text{II}}{L} = \frac{\pi^2}{3}\biggl[D_1^3 + D_2^3 +\frac{(12\Delta_1^2-1)D_1 + (12\Delta_2^2-1)D_2}{L^2} \biggr],
\end{equation}
for $N_1,N_2$ even. For an anyonic gas with fixed component densities, the degenerate phase has a higher ground state energy than the nondegenerate phase (equal only in the ferromagnetic situation). This difference persists even in the thermodynamic limit of large systems and fixed densities.

Let us further examine the features of this quantum phase transition by considering some special cases of interest. The first is the limit of a single-component anyon gas. That is, all statistical parameters are equal to each other $\kappa_{\alpha\beta} = \kappa$. In this case the degeneracy condition $\Delta_1 = \Delta_2$ is always satisfied and the ground state is described by a single Dirac sea. This case includes the mutually repulsive boson model \eqref{bosononesea} $(\kappa =0)$ and polarized pseudofermionic model $(\kappa=1/2).$

Second, we take the matrix $\bm\kappa$ to be diagonal. Here, the accumulated statistical phase $\delta_\alpha$ only depends on the number of $\alpha$-type anyons. Thus, the only coupling between components is dynamic in nature and due to hard core repulsion. The condition $\Delta_1=\Delta_2$ for a degenerate phase reduces to
\begin{equation}
	\kappa_{11}(N_1-1) - \kappa_{22}(N_2-1) =p,
\end{equation}
where $p$ is an integer. When the components have identical anyonic statistics $\kappa_{11} = \kappa_{22}$ the ground state is characterized by a single Dirac sea in the balanced case $N_1=N_2$.

Third, we consider the case in which the only nonzero elements of $\bm\kappa$ are off-diagonal. We have two mutually impenetrable boson gases, but with fractional exchange statistics between particles of different types. The ground state of the system becomes degenerate when
\begin{equation}
	\kappa_{12}(N_2-N_1) =p.
\end{equation}
This condition is fulfilled automatically in the case of a balanced mixture $N_1 = N_2$. Otherwise, degeneracy will generally not happen for irrational $\kappa_{12}$.

\begin{figure}[tb]
	\centering
		\includegraphics[width=0.95\linewidth]{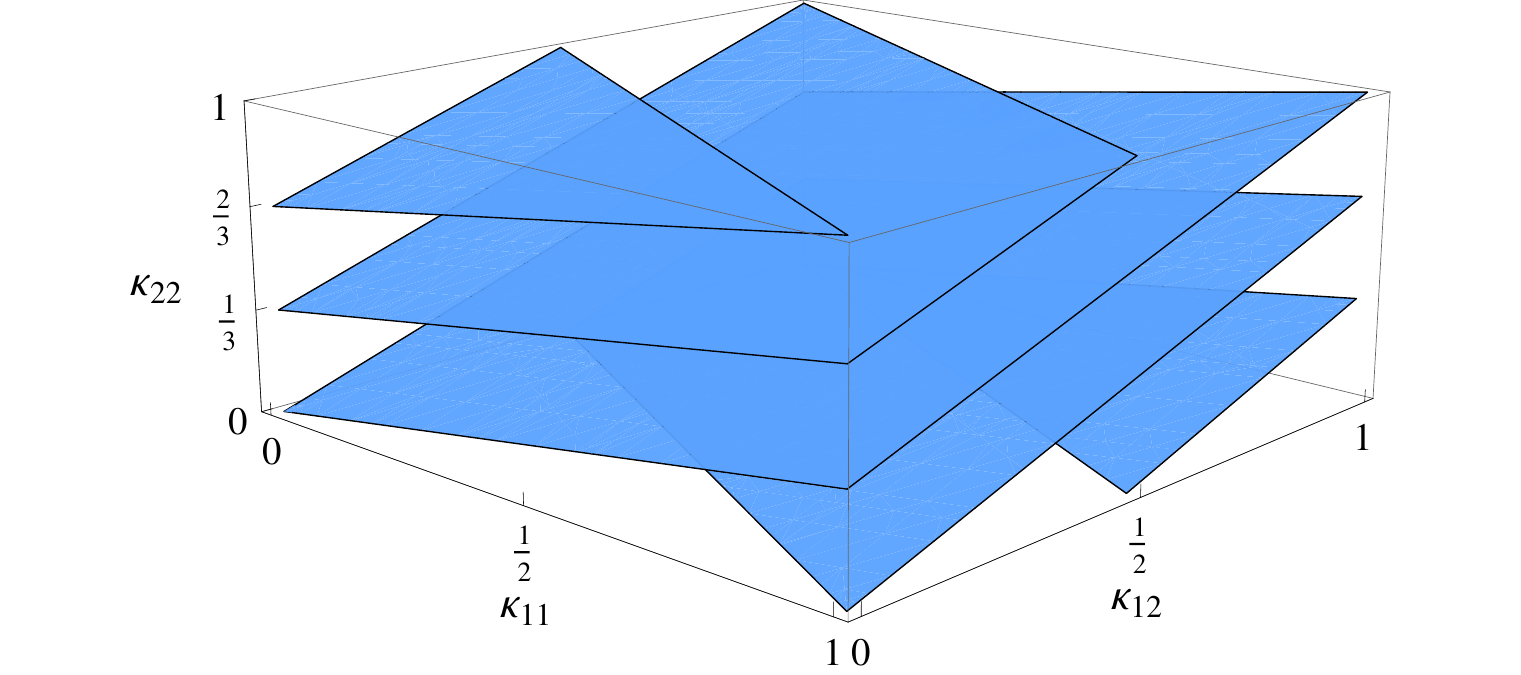}
	\caption{(Color online) Phase diagram for a gas of two type-1 and four type-2 impenetrable anyons. The (parallel) planes represent values of $\kappa_{\alpha\beta}$ for which the ground state is characterized by a single Dirac sea.}
	\label{fig:phasediagram}
\end{figure}

Next, we consider a mixture of particles and anti-particles. In this case the statistics parameters satisfy\cite{nayak2008} $\kappa_{11} = \kappa_{22}= -\kappa_{12}$. Substituting these conditions into the rule \eqref{anyonfusion} shows that an equal mixture of particles and anti-particles ($N_1=N_2$) has no overall exchange phase. This is the expected result because the vacuum obtained from pair annihilation has no phase. For an arbitrary mixture of particles and anti-particles the degeneracy condition becomes
\begin{equation}
	2\kappa_{11}(N_1-N_2)= p.
\end{equation}
Hence, the degenerate ground state always exists for the balanced case $N_1=N_2$ and also for the pseudofermionic situation $\kappa_{11} = 1/2$. 

In the most general case, the degeneracy condition $\Delta_1 = \Delta_2$ is satisfied when 
\begin{equation}\label{onesea}
	(N_1-1)\kappa_{11} +(N_2-N_1)\kappa_{12} - (N_2-1)\kappa_{22} = p.
\end{equation}
The single Dirac sea phase occurs on a family of planes in $\kappa_{\alpha\beta}$ space. Let us write the vectors $\vec{K} = (\kappa_{11},\kappa_{12},\kappa_{22})$ and $\vec{n} = (N_1-1,N_2-N_1,-N_2+1)$ in the same orthonormal basis. The condition \eqref{onesea} then becomes 
\begin{equation}\label{planeeq}
	\biggl(\vec{K} - p \frac{\vec{n}}{\vec{n}^2}\biggr)\,{\bm\cdot}\, \vec{n}= 0,
\end{equation}
which is a set of standard equations of planes labeled by the integer $p$. 
The distance $d$ between neighboring planes is \smash{$d = [(N_1-1)^2 + (N_2-N_1)^2 + (N_2-1)^2]^{-1/2}$}. Thus, in the cube $\kappa_{\alpha\beta}\in[0,1]$, the generic situation for a finite number of particles is the nondegenerate double Dirac sea phase. A sample phase diagram is given in \fig{fig:phasediagram}. 

The set of planes given by Eq.~\eqref{onesea} denote the values of the statistical parameters $\kappa_{\alpha\beta}$ and component numbers $N_\alpha$ for which the ground state is characterized by a single Dirac sea (degenerate phase). For fixed component densities, a parametric change of the statistics parameters $\kappa_{\alpha\beta}(t)$ will cross these quantum critical planes at certain points (the parameter $t$ describes a curve in $\kappa_{\alpha\beta}$ space). These quantum critical points define phases where the ground state energy is much larger than the sum of ground state energies of each component.

{\em Concluding remarks---}
We discovered that anyonic statistics in a multicomponent Tonks-Girardeau gas can lead to the formation of a new ground state with interesting physical features. The Bethe momenta in this state form two independent sets of vacancies. Each set is therefore filled independently with quasiparticles to form independent Dirac seas in the ground state. This state is remarkably different from the ground state of a mixture of impenetrable bosons, which has only one Dirac sea. For example, the formation of multiple Dirac seas significantly reduces the energy of the system and favors a balanced mixture of components (zero magnetization). This effect competes with external chemical potentials that tend to drive the system toward homogeneity (ferromagnetism). 

Moreover, the quantum phase transition described here is not observed in the single-component anyonic Lieb-Liniger model. Unlike in the single-component case, the bulk ground state energy in the multicomponent model depends significantly on the statistics parameters even for large systems.




The formalism presented here extends readily to anyonic TG gases with more than two components. In particular, the statistics-dependent phase shift $\delta_\alpha$ \eqref{statshift} generalizes to \smash{$\delta_\alpha =  - 2\pi\kappa_{\alpha\alpha} + 2\pi\sum_{\gamma=1}^{M} \kappa_{\alpha\gamma}N_\gamma$} for an $M$-component gas of impenetrable anyons.

The authors acknowledge support by the National Science Foundation through Grant Nos. DMS-0905744 and DMS-1205422. R.S. is supported by a Fulbright-CONICYT grant. 

\providecommand{\bibyu}{Yu}

\end{document}